\documentstyle[amsmath,a4,11pt]{article}

\date{}

\begin{document}

\title{{\bf Electromagnetic fields in curved spacetimes}}

\author{{Christos G
Tsagas\thanks{email:c.tsagas@damtp.cam.ac.uk}}\\ {\small DAMTP,
Centre for Mathematical Sciences, University of Cambridge}\\
{\small Wilberforce Road, Cambridge CB3 0WA}, UK}

\maketitle

\begin{abstract}
We study the evolution of electromagnetic fields in curved
spacetimes and calculate the exact wave equations of the
associated electric and magnetic components. Our analysis applies
to a general spacetime, is fully covariant and isolates all the
sources that affect the propagation of these waves. Among others,
we explicitly show how the different parts of the gravitational
field act as driving sources of electromagnetic disturbances. When
applied to perturbed FRW cosmologies, our results argue for a
superadiabatic-type amplification of large-scale cosmological
magnetic fields in Friedmann models with open spatial
curvature.
\end{abstract}

\section{Introduction}
Electromagnetic studies in curved spaces have long established the
direct coupling between the Maxwell and the Einstein fields. The
interaction emerges from the vector nature of the electromagnetic
field and from the geometrical approach to gravity introduced by
general relativity and it is interpreted as a sort of scattering
of the electromagnetic waves by the spacetime curvature.

In the present article we study electromagnetic fields in general
curved spacetimes by using the covariant approach to general
relativity. Our analysis is non-perturbative, in the sense that it
does not perturb away from a given metric but provides the full
nonlinear equations before linearising them about a chosen
background. In addition, we study the physically measurable
electric and magnetic components of the Maxwell field, rather than
the Faraday tensor or the electromagnetic 4-potential. This on one
hand complements earlier work on the subject, while on the other
it allows for a more compact mathematical presentation and for a
more transparent physical interpretation of the results. The
evolution of the Maxwell field is studied in a general spacetime
without imposing any a priori symmetries on the latter. The only
restriction is on the matter component which is of the perfect
fluid form. We derive, from first principle, the electric and
magnetic wave equations and identify all the kinematical,
dynamical and geometrical sources that drive the propagation of
these waves. We demonstrate the effects of the observers' motion
and show how the different parts of the gravitational field,
namely the Ricci and Weyl fields, affect propagating
electromagnetic disturbances. Moreover, the non-perturbative
nature of our formalism means that the nonlinear equations apply
to variety of situations of either astrophysical or cosmological
interest.

With the full equations at hand, we proceed to consider the
evolution of electromagnetic fields in spatially curved
Friedmann-Robertson-Walker (FRW) models. Noting that the
symmetries of the FRW spacetime are generally incompatible with
the generic anisotropy of the electromagnetic field, we consider
the evolution of the latter in perturbed Friedmann universes. In
particular, we look at the spacetime curvature effects on the
linear evolution of the magnetic component of the Maxwell field.
Our results show that, when the model is spatially closed, the
magnetic field has an oscillatory behaviour with a decreasing
amplitude according to the familiar $a^{-2}$-law (where $a$ is the
FRW scale-factor). The same is also true for spatially open models
with the exception of large-scale magnetic fields. There, we find
that the field decays as $a^{-1}$ and therefore that magnetic flux
conservation no longer holds at long wavelengths. This result can
be seen as an effective superadiabatic-type amplification of
large-scale magnetic fields in spatially open FRW universes due to
curvature effects alone. Crucially, the amplification is achieved
without introducing any new physics and without breaking away from
the standard properties of Maxwell's theory.

We start with an outline of the covariant formalism in section 2
and provide a covariant treatment of the electromagnetic and
gravitational fields in sections 3 and 4 respectively. The
nonlinear electromagnetic wave equations are derived in section 5
and in section 6 they are linearised and solved around curved FRW
models. We discuss our results in section 7.

\section{The covariant description}
The covariant approach to general relativity uses the kinematic
quantities of the fluid, its energy density and pressure and the
gravito-electromagnetic tensors instead of the metric which in
itself does not provide a covariant description. The key equations
are the Ricci and Bianchi identities, applied to the fluid
4-velocity vector, while Einstein's equations are incorporated via
algebraic relations between the Ricci and the energy-momentum
tensors. Here, we will only give a brief description of the
approach and direct the reader to a number of review articles for
further details and references~\cite{Eh1}-\cite{EvE}.

\subsection{The 1+3 spacetime splitting}
Consider a general spacetime with a Lorentzian metric $g_{ab}$ of
signature ($-,\,+,\,+,\,+$). Then, allow for a family of
fundamental observers living along a timelike congruence of
worldlines tangent to the 4-velocity vector
\begin{equation}
u_a=\frac{{\rm d}x^a}{{\rm d}\tau}\,,  \label{ua}
\end{equation}
where $\tau$ is the associated proper time and
$u_au^a=-1$~\cite{EvE}. This fundamental velocity field introduces
an local, 1+3 ``threading'' of the spacetime into time and space.
The vector $u_a$ determines the time direction and the tensor
$h_{ab}=g_{ab}+u_au_b$ projects orthogonal to $u_a$ into what is
known as the observers' instantaneous rest space. Note that, in
the absence of rotation, $h_{ab}$ also acts as the metric of the
spatial sections.

Employing $u_a$ and $h_{ab}$ one defines the covariant time
derivative and the orthogonally projected gradient of any given
tensor field $T_{ab\cdots}{}^{cd\cdots}$ according to
\begin{equation}
\dot{T}_{ab\cdots}{}^{cd\cdots}=
u^e\nabla_eT_{ab\cdots}{}^{cd\cdots} \hspace{5mm} {\rm and}
\hspace{5mm} {\rm D}_eT_{ab\cdots}{}^{cd\cdots}=
h_e{}^sh_a{}^fh_b{}^ph_q{}^ch_r{}^d\cdots\nabla_s
T_{fp\cdots}{}^{qr\cdots}\,, \label{deriv}
\end{equation}
respectively. The former indicates differentiation along the
timelike direction and the latter operates on the observers' rest
space.

\subsection{The matter field}
Relative to the aforementioned fundamental observers, the
energy-momentum tensor of a general imperfect fluid decomposes
into its irreducible parts as~\cite{EvE}\footnote{Throughout this
article we use geometrised units with $c=1=8\pi G$. Consequently,
all geometrical variables have physical dimensions that are
integer powers of length.}
\begin{equation}
T_{ab}=\mu u_au_b+ ph_{ab}+ 2q_{(a}u_{b)}+ \pi_{ab}\,.
\label{Tab}
\end{equation}
Here, $\mu=T_{ab}u^au^b$ and $p=T_{ab}h^{ab}/3$ are respectively
the energy density and the isotropic pressure of the medium,
$q_a=-h_a{}^bT_{bc}u^c$ is the energy-flux vector relative to
$u_a$ and $\pi_{ab}=h_{\langle a}{}^ch_{b\rangle}{}^dT_{cd}$ is
the symmetric and trace-free tensor that describes the anisotropic
pressure of the fluid.\footnote{Angled brackets denote the
symmetric and trace-free part of projected second-rank tensors and
the orthogonally projected component of vectors.} It follows that
$q_au^a=0=\pi_{ab}u^a$. When the fluid is perfect both $q_a$ and
$\pi_{ab}$ are identically zero and the remaining degrees of
freedom are determined by the equation of state. For a barotropic
medium the latter reduces to $p=p(\mu)$, with $c_{\rm s}^2={\rm
d}p/{\rm d}\mu$ representing the associated adiabatic sound speed.

When dealing with a multi-component medium, one needs to account
for the velocity ``tilt'' between the various matter components
and the fundamental observers~\cite{KE}. Here, however, we will
consider a single-component fluid and we will assume that the
fundamental observers are moving along with it.

\subsection{The covariant kinematics}
The observers' motion is characterised by the irreducible
kinematical quantities of the $u_a$-congruence, which emerge from
the following covariant decomposition of the 4-velocity gradient
\begin{equation}
\nabla_bu_a=\sigma_{ab}+ \omega_{ab}+ {\textstyle{1\over3}}\Theta
h_{ab}- \dot{u}_au_b\,,  \label{Nbua}
\end{equation}
where $\sigma_{ab}={\rm D}_{\langle b}u_{a\rangle}$,
$\omega_{ab}={\rm D}_{[b}u_{a]}$, $\Theta=\nabla^au_a={\rm
D}^au_a$ and $\dot{u}_a=u^b\nabla_bu_a$ are respectively the shear
and the vorticity tensors, the expansion (or contraction) scalar
and the 4-acceleration vector~\cite{EvE}. Then,
$\sigma_{ab}u^a=0=\omega_{ab}u^a=\dot{u}_au^a$ by definition.
Also, on using the orthogonally projected alternating tensor
$\epsilon_{abc}$ (with
$\dot{\epsilon}_{abc}=3u_{[a}\epsilon_{bc]d}\dot{u}^d$), one
defines the vorticity vector
$\omega_a=\epsilon_{abc}\omega^{bc}/2$.

The nonlinear covariant kinematics are determined by a set of
three propagation equations complemented by an equal number of
constraints~\cite{EvE}. The former contains Raychaudhuri's formula
\begin{equation}
\dot{\Theta}=-{\textstyle{1\over3}}\Theta^2-
{\textstyle{1\over2}}(\mu+3p)- 2(\sigma^2-\omega^2)+ {\rm
D}^a\dot{u}_a+ \dot{u}_a\dot{u}^a\,,  \label{Ray}
\end{equation}
for the time evolution of $\Theta$, the shear propagation equation
\begin{equation}
\dot{\sigma}_{\langle ab\rangle}=
-{\textstyle{2\over3}}\Theta\sigma_{ab}- \sigma_{c\langle
a}\sigma^c{}_{b\rangle}- \omega_{\langle a}\omega_{b\rangle}+ {\rm
D}_{\langle a}\dot{u}_{b\rangle}+ \dot{u}_{\langle
a}\dot{u}_{b\rangle}- E_{ab}+ {\textstyle{1\over2}}\pi_{ab}\,,
\label{sigmadot}
\end{equation}
which describes kinematical anisotropies, and the evolution
equation of the vorticity
\begin{equation}
\dot{\omega}_{\langle a\rangle}=
-{\textstyle{2\over3}}\Theta\omega_a- {\textstyle{1\over2}}{\rm
curl}\dot{u}_a+ \sigma_{ab}\omega^b\,.  \label{omegadot}
\end{equation}
Note that $\sigma^2=\sigma_{ab}\sigma^{ab}/2$ and
$\omega^2=\omega_{ab}\omega^{ab}/2=\omega_a\omega^a$ are
respectively the magnitudes of the shear and the vorticity, while
$E_{ab}$ is the electric component of the Weyl tensor (see section
3.2 below). Also, ${\rm curl}v_a=\epsilon_{abc}{\rm D}^bv^c$ for
any orthogonally projected vector $v_a$ by definition.

Equations (\ref{Ray}), (\ref{sigmadot}) and (\ref{omegadot}) are
complemented by a set of three nonlinear constraints. These are
the shear
\begin{equation}
{\rm D}^b\sigma_{ab}={\textstyle{2\over3}}{\rm D}_a\Theta+ {\rm
curl}\omega_a+ 2\epsilon_{abc}\dot{u}^b\omega^c- q_a\,,
\label{shearcon}
\end{equation}
the vorticity
\begin{equation}
{\rm D}^a\omega_a=\dot{u}_a\omega^a\,,  \label{omegacon}
\end{equation}
and the magnetic Weyl constraint
\begin{equation}
H_{ab}={\rm curl}\sigma_{ab}+ {\rm D}_{\langle
a}\omega_{b\rangle}+ 2\dot{u}_{\langle a}\omega_{b\rangle}\,,
\label{Hcon}
\end{equation}
where ${\rm curl}T_{ab}=\epsilon_{cd\langle a}{\rm
D}^cT_{b\rangle}{}^d$ for any orthogonally projected tensor
$T_{ab}$.

\section{The electromagnetic field}
Covariant studies of electromagnetic fields date back to the work
of Ehlers~\cite{Eh1} and Ellis~\cite{E2}. In addition to its
inherent mathematical compactness and clarity, the covariant
formalism facilitates a physically intuitive fluid description of
the Maxwell field. In particular, the latter is represented as an
imperfect fluid with properties specified by its electric and
magnetic components.

\subsection{The electric and magnetic components}
The Maxwell field is covariantly characterised by the
antisymmetric electromagnetic (Faraday) tensor $F_{ab}$, which
relative to a fundamental observer decomposes into an electric and
a magnetic component as~\cite{E2,TB1}
\begin{equation}
F_{ab}=2u_{[a}E_{b]}+ \epsilon_{abc}H^c\,.  \label{Fab}
\end{equation}
In the above $E_a=F_{ab}u^b$ and $H_a=\epsilon_{abc}F^{bc}/2$ are
respectively the electric and magnetic fields experienced by the
observer. Note that $E_au^a=0=H_au^a$, ensuring that both $E_a$
and $H_a$ are spacelike vectors living in the observer's
3-dimensional rest-space. Also, expression
$H_a=\epsilon_{abc}F^{bc}/2$ guarantees that $H_a$ is the dual of
the antisymmetric (pseudo) tensor $F_{ab}$.

The Faraday tensor also determines the energy-momentum tensor of
the Maxwell field. In particular we have
\begin{equation}
T^{(em)}_{ab}=-F_{ac}F^c{}_b-
{\textstyle{1\over4}}F_{cd}F^{cd}g_{ab}\,,  \label{Tem1}
\end{equation}
which, on using (\ref{Fab}), provides an irreducible decomposition
for $T^{(em)}_{ab}$. More precisely, relative to a fundamental
observer, the latter splits into~\cite{E2,TB1}
\begin{eqnarray}
T^{(em)}_{ab}={\textstyle{1\over2}}(E^2+H^2)u_au_b+
{\textstyle{1\over6}}(E^2+H^2)h_{ab}+ 2{\cal Q}_{(a}u_{b)}+ {\cal
P}_{ab}\,.  \label{Tem}
\end{eqnarray}
Here $E^2=E_aE^a$ and $H^2=H_aH^a$ are the magnitudes of the two
fields, ${\cal Q}_a=\epsilon_{abc}E^bH^c$ is the electromagnetic
Poynting vector and ${\cal P}_{ab}$ is a symmetric, trace-free
tensor given by
\begin{equation}
{\cal P}_{ab} ={\cal P}_{\langle ab\rangle}
={\textstyle{1\over3}}(E^2+H^2)h_{ab}- E_aE_b- H_aH_b\,.
\label{cP}
\end{equation}
Expression (\ref{Tem}) provides a fluid description of the Maxwell
field and manifests its generically anisotropic nature. In
particular, the electromagnetic field corresponds to an imperfect
fluid with energy density $(E^2+H^2)/2$, isotropic pressure
$(E^2+H^2)/6$, anisotropic stresses given by ${\cal P}_{ab}$ and
an energy-flux vector represented by ${\cal Q}_a$. Equation
(\ref{Tem}) also ensures that $T_a^{(em)\;a}=0$, in agreement with
the trace-free nature of the radiation stress-energy tensor.
Finally, we note that by putting the isotropic and anisotropic
pressure together one arrives at the familiar Maxwell tensor,
which assumes the covariant form
\begin{equation}
{\cal M}_{ab}={\textstyle{1\over2}}(E^2+H^2)h_{ab}- E_aE_b-
H_aH_b\,.  \label{cM}
\end{equation}

\subsection{Maxwell's equations}
We follow the evolution of the electromagnetic field by means of
Maxwell's equations. In their standard tensor form the latter read
\begin{equation}
\nabla_{[c}F_{ab]}=0\,, \hspace{15mm}{\rm and}\hspace{15mm}
\nabla^bF_{ab}=J_a\,,  \label{Max}
\end{equation}
where (\ref{Max}a) manifests the existence of a 4-potential and
$J_a$ is the 4-current that sources the electromagnetic field,
With respect to the $u_a$-congruence, the 4-current splits into
its irreducible parts according to
\begin{equation}
J_a=\rho_{\em e}u_a+ {\cal J}_a\,,  \label{Ja}
\end{equation}
with $\rho_{\em e}=-J_au^a$ representing the charge density and
${\cal J}_a=h_a{}^bJ_b$ the orthogonally projected current
(i.e.~${\cal J}_au^a=0$).

Relative to a fundamental observer, each one of Maxwell's
equations decomposes into a timelike and a spacelike component.
Thus, by projecting (\ref{Max}a) and (\ref{Max}b) along and
orthogonal to the 4-velocity vector $u_a$, we obtain a set of two
propagation equations~\cite{E2,TB1}
\begin{equation}
\dot{E}_{\langle a\rangle}=
\left(\sigma_{ab}+\varepsilon_{abc}\omega^c
-{\textstyle{2\over3}}\Theta h_{ab}\right)E^b+
\varepsilon_{abc}\dot{u}^bH^c+ {\rm curl}H_a- {\cal J}_a\,,
\label{M1}
\end{equation}
\begin{equation}
\dot{H}_{\langle a\rangle}=
\left(\sigma_{ab}+\varepsilon_{abc}\omega^c
-{\textstyle{2\over3}}\Theta h_{ab}\right)H^b-
\varepsilon_{abc}\dot{u}^bE^c- {\rm curl}E_a\,, \label{M2}
\end{equation}
and the following pair of constraints
\begin{equation}
{\rm D}^aE_a+ 2\omega^aH_a=\rho_{\rm e}\,,  \label{M3}
\end{equation}
\begin{equation}
{\rm D}^aH_a- 2\omega^aE_a=0\,.  \label{M4}
\end{equation}
Note that in addition to the usual ``curl'' and ``divergence''
terms, there are terms due to the observer's motion. According to
Eq.~(\ref{M3}), in the absence of an electric field the observed
charge density is $\rho_{\rm e}= 2\omega^aH_a$. This means nonzero
charge density unless $\omega^aH_a=0$ (see~\cite{CBF} for a
discussion on the charge asymmetry of the universe). Also,
following (\ref{M4}), the magnetic vector is not a solenoidal
unless $\omega^aE_a=0$.

\subsection{The conservation laws}
The antisymmetry of the Faraday tensor (see Eq.~(\ref{Fab})) and
the second of Maxwell's formulae (see Eq.~(\ref{Max}b)) imply the
conservation law
\begin{equation}
\nabla^aJ_a=0\,,  \label{ccon}
\end{equation}
for the 4-current density. Then, on using decomposition
(\ref{Ja}), expression (\ref{ccon}) provides the covariant form of
the charge density conservation law~\cite{E2,T}
\begin{equation}
\dot{\rho}_{\rm e}=-\Theta\rho_{\rm e}- {\rm D}^a{\cal J}_a-
\dot{u}^a{\cal J}_a\,.  \label{chcon}
\end{equation}
Thus, in the absence of spatial currents, the charge density
evolution depends entirely on the average volume expansion (or
contraction) of the fluid element.

\subsection{Ohm's law}
The electrical conductivity of the medium determines the relation
between the 4-current and the associated electric field via Ohm's
law. In covariant form the latter reads
\begin{equation}
J_a-\rho_{\em e}u_a=\sigma E_a\,,  \label{Ohm}
\end{equation}
where $\sigma$ is the scalar conductivity of the medium~\cite{J}.
Projecting the above into the observer's rest space one arrives at
\begin{equation}
{\cal J}=\sigma E_a\,.  \label{Ohm1}
\end{equation}
Thus, non-zero spatial currents are compatible with a vanishing
electric field as long as the conductivity of the medium is
infinite (i.e.~for $\sigma\rightarrow\infty$). Alternatively, one
can say that at the infinite conductivity limit, which defines the
well known MHD approximation, the electric field vanishes in the
frame of the fluid. On the other hand, zero electrical
conductivity implies that the spatial currents vanish even when
the electric field is non-zero.

\section{The gravitational field}
Covariantly, the local gravitational field is monitored by a set
of algebraic relations between the Ricci curvature tensor and the
energy-momentum tensor of the matter. The free gravitational
field, on the other hand, is described by the electric and
magnetic components of the conformal curvature (Weyl) tensor.

\subsection{The local Ricci curvature}
In the general relativistic geometrical interpretation of gravity,
matter determines the spacetime curvature which in turn dictates
the motion of the matter. This interaction is manifested in the
Einstein field equations, which in the absence of a cosmological
constant take the form
\begin{equation}
R_{ab}=T_{ab}- {\textstyle{1\over2}}Tg_{ab}\,,  \label{EFE}
\end{equation}
where $R_{ab}=R_{acb}{}^c$ is the spacetime Ricci tensor, $T_{ab}$
is the energy-momentum tensor of the matter fields, with
$T=T_a{}^a$ being the trace. For our purposes the total
energy-momentum tensor has the form
$T_{ab}=T^{(f)}_{ab}+T^{(em)}_{ab}$, where $T^{(f)}_{ab}$ is given
by Eq.~(\ref{Tab}) and $T^{(em)}_{ab}$ by Eq.~(\ref{Tem}). Thus,
\begin{equation}
T_{ab}=\left[\mu+{\textstyle{1\over2}}(H^2+E^2)\right]u_au_b+
\left[p+{\textstyle{1\over6}}(H^2+E^2)\right]h_{ab}+
2(q_{(a}+{\cal Q}_{(a})u_{b)}+ \pi_{ab}+ {\cal P}_{ab}\,,
\label{tTab}
\end{equation}
ensuring that $\mu+(H^2+E^2)/2$ is the total energy density of the
system, $p+(H^2+E^2)/6$ is the total isotropic pressure,
$q_a+{\cal Q}_a$ is the total heat flux vector and $\pi_{ab}+{\cal
P}_{ab}$ is the total anisotropic pressure. The inclusion of
electromagnetic terms in the energy-momentum tensor of the matter
guarantees that the contribution of the Maxwell field on the
spacetime geometry is fully accounted for.

Starting from the Einstein field equations and assuming that
$T_{ab}$ is given by Eq.~(\ref{tTab}), we arrive at the following
algebraic relations~\cite{T}
\begin{eqnarray}
R_{ab}u^au^b&=&{\textstyle{1\over2}}(\mu+3p+E^2+H^2)\,,
\label{EFE1}\\ h_a{}^bR_{bc}u^c&=&-(q_a+{\cal Q}_a)\,,
\label{EFE2}\\ h_a{}^ch_b{}^dR_{cd}&=&\left[{\textstyle{1\over2}}
\left(\mu-p+{\textstyle{1\over3}}E^2
+{\textstyle{1\over3}}H^2\right)\right]h_{ab}+ \pi_{ab}+ {\cal
P}_{ab}\,.  \label{EFE3}
\end{eqnarray}
In addition, the trace of (\ref{EFE}) gives $R=-T$, with
$R=R_a{}^a$ and $T=T_a{}^a=3p-\mu$, where the latter result is
guaranteed by the trace-free nature of $T^{(em)}_{ab}$. Note that
the above expressions are valid irrespective of the strength of
the electromagnetic components. When the Maxwell field is weak
relative to the matter, namely for $E^2,\,H^2\ll\mu$, one might
treat the electromagnetic contribution to the spacetime curvature
as a first order perturbation. Finally, recall that
$q_a=0=\pi_{ab}$ when dealing with a perfect fluid.

\subsection{The long-range Weyl curvature}
The Ricci tensor describes the local gravitational field of the
nearby matter. The long-range gravitational field, namely
gravitational waves and tidal forces, propagates through the Weyl
conformal curvature tensor. The splitting of the gravitational
field into its local and non-local components is demonstrated in
the following decomposition of the Riemann tensor
\begin{equation}
R_{abcd}=C_{abcd}+
{\textstyle{1\over2}}\left(g_{ac}R_{bd}+g_{bd}R_{ac}-g_{bc}R_{ad}
-g_{ad}R_{bc}\right)-
{\textstyle{1\over6}}R\left(g_{ac}g_{bd}-g_{ad}g_{bc}\right)\,,
\label{Riemann}
\end{equation}
where $C_{abcd}$ is the Weyl tensor. The latter shares all the
symmetries of the Riemann tensor and is also trace-free
(i.e.~$C^c{}_{acb}=0$). Relative to the fundamental observers, the
Weyl tensor decomposes further into its irreducible parts
according to
\begin{equation}
C_{abcd}=
\left(g_{abqp}g_{cdsr}-\eta_{abqp}\eta_{cdsr}\right)u^qu^sE^{pr}-
\left(\eta_{abqp}g_{cdsr}+g_{abqp}\eta_{cdsr}\right)u^qu^sH^{pr}\,,
\label{Weyl}
\end{equation}
where $g_{abcd}=g_{ac}g_{bd}-g_{ad}g_{bc}$ (e.g.~see~\cite{HE,M}).
The symmetric and trace-free tensors $E_{ab}$ and $H_{ab}$ are
known as the electric and magnetic Weyl components and they are
given by
\begin{equation}
E_{ab}=C_{acbd}u^cu^d\,, \hspace{15mm}{\rm and}\hspace{15mm}
H_{ab}={\textstyle{1\over2}}\epsilon_a{}^{cd}C_{cdbe}u^e\,,
\label{EHab}
\end{equation}
with $E_{ab}u^b=0=H_{ab}u^b$. Given that $E_{ab}$ has a Newtonian
counterpart, the electric part of the Weyl tensor is associated
with the tidal field. The magnetic component, on the other hand,
has no Newtonian analogue and therefore is primarily associated
with gravitational waves~\cite{E1}. Of course, both tensors are
required if gravitational waves are to exist. For a comparison
with the non-covariant metric based treatments of gravitational
waves we note that in perturbed FRW models the harmonically
decomposed, pure-tensor metric perturbation is
$H_{T}=2E+\sigma^{\prime}/{\rm n}$~\cite{L}. Here, $E$ and
$\sigma$ represent the harmonic parts of the transverse traceless
electric Weyl and shear tensors respectively. Also, ${\rm n}$ is
the associated wavenumber and a prime denotes derivatives with
respect to conformal time.

The Weyl tensor represents the part of the curvature that is not
determined locally by matter. However, the dynamics of the Weyl
field is not entirely arbitrary because the Riemann tensor
satisfies the Bianchi identities. When contracted the latter take
the form~\cite{HE}
\begin{equation}
\nabla^dC_{abcd}=\nabla_{[b}R_{a]c}+
{\textstyle{1\over6}}\,g_{c[b}\nabla_{a]}R\,,  \label{Bianchi}
\end{equation}
by means of decomposition (\ref{Riemann}). In a sense the
contracted Bianchi identities act as the field equations for the
Weyl tensor, determining the part of the spacetime curvature that
depends on the matter distribution at other points~\cite{HE}. The
form of the contracted Bianchi identities guarantees that once the
electromagnetic contribution to the Ricci curvature has been
incorporated, through the Einstein field equations, the effect of
the Maxwell field on the Weyl curvature has also been fully
accounted for.

Expression (\ref{Bianchi}) splits into a set of two propagation
and two constraint equations, which monitor the evolution of the
electric and magnetic Weyl components~\cite{E1}-\cite{EvE}. These
formulae are not used to derive the electromagnetic wave equations
of section 5.3 and are therefore not essential for our purposes.
Here we simply note that the aforementioned set of equations is
remarkably similar to Maxwell's formulae, which in turn explains
the names of $E_{ab}$ and $H_{ab}$. This Maxwell-like form of the
free gravitational field underlines the rich correspondence
between electromagnetism and general relativity, which has been
the subject of theoretical debate for many decades
(see~\cite{B}-\cite{Da} for a representative list).

\section{The electromagnetic wave equations}
Studies of electromagnetic waves in curved spacetimes have long
established that, while propagating similar to any other
travelling wave, electromagnetic disturbances also interact with
the spacetime curvature. As a result, electromagnetic signals
propagate inside as well as on the future light cone of an event,
indicating the failure of Huygens' principle in curved
spaces~\cite{DWB}-\cite{G}.

\subsection{The wave equation for the electromagnetic field
tensor}
Maxwell's equations immediately provide a wave equation for the
electromagnetic field tensor. In particular, taking the covariant
derivative of (\ref{Max}a) and using (\ref{Max}b) we arrive at
\begin{equation}
\nabla^2F_{ab}=-2R_{acbd}F^{cd}+ R_a{}^cF_{cb}+
F_a{}^cR_{cb}+ \nabla_bJ_a- \nabla_aJ_b\,,  \label{wFab}
\end{equation}
where $\nabla^2=\nabla^a\nabla_a$ is the generalised covariant
Laplacian operator (e.g.~see~\cite{Eh2,N}). The above, which holds
in a general spacetime, reveals the role of the curvature as a
driving source of electromagnetic disturbances. Note that the
Riemann and Ricci curvature terms in the left-hand side of
Eq.~(\ref{wFab}) emerge after using the Ricci identity
\begin{equation}
2\nabla_{[a}\nabla_{b]}F_{cd}=R_{abce}F^e{}_d+ R_{abde}F_c{}^e\,,
\label{RicciFab}
\end{equation}
which here monitors the commutation between the covariant
derivatives of $F_{ab}$. Expression (\ref{wFab}) can also provide
the individual wave equations for the electric and magnetic
components of $F_{ab}$. For example, contracting Eq.~(\ref{wFab})
along $u_a$ eventually leads to the wave equation of $E_a$, while
its dual provides the magnetic wave equation. Here, we will follow
an alternative route and obtain these expressions directly from
the decomposed Maxwell formulae (\ref{M1}) and (\ref{M2}).

\subsection{The electro/magneto-curvature coupling}
In addition to the Einstein field equations, vector sources, like
the electromagnetic field, obey an extra set of equations, known
as the Ricci identities, which manifest the direct interaction
between electromagnetism and spacetime geometry. This coupling
emerges naturally from the vector nature of the Maxwell field and
from the geometrical approach to gravity of general relativity.
When applied to the magnetic field vector the Ricci identity reads
\begin{equation}
2\nabla_{[c}\nabla_{b]}H_a=R_{dabc}H^d\,;
\label{Ricci}
\end{equation}
with an exactly analogous expression for the electric component.
Clearly, on using decomposition (\ref{Riemann}), the Ricci
identity couples the electromagnetic field explicitly with both
the local and the long-range gravitational field. Also, by
projecting the above into the observer's rest space one arrives at
what is known as the 3-Ricci identity
\begin{equation}
2{\rm D}_{[c}{\rm D}_{b]}H_a=
-2\varepsilon_{cbd}\omega^d\dot{H}_{\langle a\rangle}+
{\cal R}_{dabc}H^d\,,  \label{3Ricci}
\end{equation}
describing the interaction between the magnetic field and the
local spatial geometry~\cite{TB2,TM1}. Clearly an exactly
analogous relation holds for $E_a$ as well. Note that ${\cal
R}_{abcd}$ is the orthogonally projected part of $R_{abcd}$,
namely the Riemann tensor of the observer's local 3-space. Note
that the validity of both (\ref{Ricci}) and (\ref{3Ricci}) extends
to any arbitrary spacetime (e.g.~see~\cite{E1,HE}).

\subsection{The wave equations for the electric and magnetic
fields}
Equations (\ref{M1})-(\ref{M2}) monitor the propagation of
electromagnetic fields in a general spacetime either in vacuum
(i.e.~for source-free fields with $\rho_{\rm e}=0={\cal J}_a$) or
in the presence of matter. Starting form these formulae one can
work out the wave equations for propagating electromagnetic
radiation in a general spacetime. In particular, taking the time
derivative of Eq.~(\ref{M1}) one obtains the wave-like evolution
equation of the electric field. Similarly, the time derivative of
Eq.~(\ref{M2}) leads to the corresponding wave equation of the
magnetic field. In the Minkowski space of special relativity these
calculations are relatively straightforward since the geometry of
the space is trivial. In the context of general relativity,
however, this is no longer true and one has to account for the
coupling between the electro-magnetic fields and the spacetime
geometry discussed earlier. Technically speaking, this requires
using the Ricci identities and leads to spatial curvature terms
every time the projected derivatives of $E_a$ or $H_a$ commute. In
addition, the Ricci identities guarantee a Weyl field contribution
whenever a time derivative and a projected gradient of either the
electric or the magnetic field commute.

Assuming that the matter component has a perfect fluid form with a
barotropic equation of state, we take the time derivative of
Eq.~(\ref{M1}) and project it orthogonal to $u_a$. Then, using the
kinematical propagation and constraint equations of section 2.3,
expression (\ref{cP}), relations (\ref{M2})-(\ref{Weyl}) and the
commutation laws (\ref{Ricci}), (\ref{3Ricci}) we arrive at the
following wave equation for the electric field vector
\begin{eqnarray}
\ddot{E}_{\langle a\rangle}-{\rm D}^2E_a&=&
{\textstyle{1\over3}}\mu(1+3w)E_a+
\left(\sigma_{ab}-\varepsilon_{abc}\omega^c
-{\textstyle{5\over3}}\Theta h_{ab}\right)\dot{E}^b+
{\textstyle{1\over3}}\Theta\left(\sigma_{ab}
+\varepsilon_{abc}\omega^c -{\textstyle{4\over3}}\Theta
h_{ab}\right)E^b\nonumber\\&{}& -\sigma_{\langle
a}{}^c\sigma_{b\rangle c}E^b+
\varepsilon_{abc}E^b\sigma^{cd}\omega_d+ {\textstyle{4\over3}}
\left(\sigma^2-{\textstyle{2\over3}}\omega^2\right)E_a+
{\textstyle{1\over3}}\omega_{\langle a}\omega_{b\rangle}E^b+
\dot{u}^b\dot{u}_bE_a\nonumber\\&{}&
-{\textstyle{5\over2}}\varepsilon_{abc}\dot{u}^b{\rm curl}E^c+
{\rm D}_{\langle a}E_{b\rangle}\dot{u}^b+
{\textstyle{2\over3}}\varepsilon_{abc}H^b{\rm D}^c\Theta+
\varepsilon_{abc}H_d{\rm D}^b\sigma^{cd}+ {\rm D}_{\langle
a}\omega_{b\rangle}H^b\nonumber\\&{}&+
{\textstyle{3\over2}}\varepsilon_{abc}H^b{\rm curl}\omega^c+ 2{\rm
D}_{\langle a}H_{b\rangle}\omega^b-
2\varepsilon_{abc}\sigma^b{}_d{\rm D}^{\langle c}H^{d\rangle}+
\varepsilon_{abc}\ddot{u}^bH^c+
{\textstyle{7\over3}}\dot{u}^b\omega_bH_a\nonumber\\&{}&+
{\textstyle{4\over3}}H^b\omega_b\dot{u}_a- 3\dot{u}^bH_b\omega_a+
3\varepsilon_{abc}\dot{u}^b\sigma^{cd}H_d+
{\textstyle{1\over3}}\rho_{\rm e}\dot{u}_a- {\rm D}_a\rho_{\rm
e}-\Theta{\cal J}_a- \dot{{\cal J}}_a\nonumber\\&{}& -{\cal
R}_{ab}E^b- E_{ab}E^b+ H_{ab}H^b\,. \label{ddotEa}
\end{eqnarray}
Similarly, one may start from Eq.~(\ref{M2}) and proceed in an
analogous way to obtain the wave equation of the magnetic field
vector
\begin{eqnarray}
\ddot{H}_{\langle a\rangle}-{\rm D}^2H_a&=&
{\textstyle{1\over3}}\mu(1+3w)H_a+
\left(\sigma_{ab}-\varepsilon_{abc}\omega^c
-{\textstyle{5\over3}}\Theta h_{ab}\right)\dot{H}^b+
{\textstyle{1\over3}}\Theta\left(\sigma_{ab}
+\varepsilon_{abc}\omega^c -{\textstyle{4\over3}}\Theta
h_{ab}\right)H^b\nonumber\\&{}& -\sigma_{\langle
a}{}^c\sigma_{b\rangle c}H^b+
\varepsilon_{abc}H^b\sigma^{cd}\omega_d+
{\textstyle{4\over3}}\left(\sigma^2
-{\textstyle{2\over3}}\omega^2\right)H_a+
{\textstyle{1\over3}}\omega_{\langle a}\omega_{b\rangle}H^b+
\dot{u}^b\dot{u}_bH_a\nonumber\\&{}&
-{\textstyle{5\over2}}\varepsilon_{abc}\dot{u}^b{\rm curl}H^c+
{\rm D}_{\langle a}H_{b\rangle}\dot{u}^b-
{\textstyle{2\over3}}\varepsilon_{abc}E^b{\rm D}^c\Theta-
\varepsilon_{abc}E_d{\rm D}^b\sigma^{cd}- {\rm D}_{\langle
a}\omega_{b\rangle}E^b\nonumber\\&{}&
-{\textstyle{3\over2}}\varepsilon_{abc}E^b{\rm curl}\omega^c-
2{\rm D}_{\langle a}E_{b\rangle}\omega^b+
2\varepsilon_{abc}\sigma^b{}_d{\rm D}^{\langle c}E^{d\rangle}-
\varepsilon_{abc}\ddot{u}^bE^c-
{\textstyle{7\over3}}\dot{u}^b\omega_bE_a\nonumber\\&{}&
-{\textstyle{4\over3}}E^b\omega_b\dot{u}_a+ 3\dot{u}^bE_b\omega_a-
3\varepsilon_{abc}\dot{u}^b\sigma^{cd}E_d-
{\textstyle{2\over3}}\rho_{\rm e}\omega_a+
2\varepsilon_{abc}\dot{u}^b{\cal J}^c+{\rm curl}{\cal
J}_a\nonumber\\&{}& -{\cal R}_{ab}H^b- E_{ab}H^b- H_{ab}E^b\,.
\label{ddotHa}
\end{eqnarray}
As expected, when there are no charges and currents, one recovers
Eq.~(\ref{ddotHa}) from (\ref{ddotEa}) by simply replacing $E_a$
with $H_a$ and $H_a$ with $-E_a$. Similarly, we obtain
(\ref{ddotEa}) from (\ref{ddotHa}) after substituting $H_a$ with
$E_a$ and $E_a$ with $-H_a$. In the presence of charges and
currents, however, this symmetry no longer holds and the apparent
breakdown reflects the absence of magnetic monopoles.

The above expressions provide a covariant description of
propagating electromagnetic waves in a general spacetime and
incorporate the electromagnetic input to the curvature of the
latter.\footnote{By including the Maxwell field in the Einstein
field equations (see Eqs.~(\ref{EFE})-(\ref{EFE3})) the
electromagnetic contribution to the spacetime geometry has been
fully accounted for. In practise this means ensuring that $\mu$
has been replaced with $\mu+(E^2+H^2)/2$, $p$ with
$p+(E^2+H^2)/6$, $q_a$ with ${\cal Q}_a$ and $\pi_{ab}$ with
${\cal P}_{ab}$ in every formula used to derive
Eqs.~(\ref{ddotEa}) and (\ref{ddotHa}). For example, by
implementing the aforementioned substitution into the kinematical
expressions of section 2.3, we incorporate fully the
electromagnetic impact on the model's kinematics.}. So far the
only restrictions are those imposed on the fluid, which has a
barotropic equation of state. That aside, Eqs.~(\ref{ddotEa}) and
(\ref{ddotHa}) are fully nonlinear in perturbative terms. Once the
background is specified, these equations can describe the
evolution of the electromagnetic field at any perturbative level.
In general, of course, one needs to couple these formulae with the
appropriate propagation equations of the various kinematical,
dynamical and geometrical variables that appear in the right-hand
side of (\ref{ddotEa}) and (\ref{ddotHa}). Clearly, the more
complicated the background the more equations are necessary for
the system to close.

Among others, the above given wave equations show how the
kinematical quantities, namely the expansion, the shear, the
vorticity and the acceleration drive the propagation of
electromagnetic waves. Here, the barotropic nature of the matter
component means that the 4-acceleration takes the form
\begin{equation}
\mu(1+w)\dot{u}_a=-{\rm D}_ap+ \rho_{\rm
e}E_a+\varepsilon_{abc}{\cal J}^bH^c\,,  \label{accel}
\end{equation}
with  contributions from gradients in the fluid pressure and from
the electromagnetic Lorentz force only. The input from the
spacetime geometry to Eqs.~(\ref{ddotEa}) and (\ref{ddotHa}) is
through the spatial and the Weyl curvature components. The former
is represented by ${\cal R}_{ab}$, the orthogonally projected
3-Ricci tensor, defined by
\begin{equation}
{\cal R}_{ab} ={\cal R}^c{}_{acb} =h_a{}^ch_b{}^dR_{cd}+
R_{acbd}u^cu^d+ v_{ac}v^c{}_b- \Theta v_{ab}\,,  \label{3Rab}
\end{equation}
where $v_{ab}={\rm D}_bu_a$ is the second fundamental form
describing the extrinsic curvature of the space
(e.g.~see~\cite{HE,TM1}). Note that the tidal part of the Weyl
field contributes to the evolution of either $E_a$ or $H_a$ via
its direct coupling with the aforementioned fields. The effect of
the magnetic Weyl tensor, on the other hand, is indirect and
requires the presence of both the electromagnetic field
components.

The non-perturbative nature of our analysis, namely the fact that
we have not yet specified our background spacetime, means that
Eqs.~(\ref{ddotEa}) and (\ref{ddotHa}) apply to a range of
physical situations (e.g.~see~\cite{Ge}-\cite{KM}). For example,
in the absence of matter sources one can always set the observer's
acceleration to zero (see Eq.~(\ref{accel})). If, in addition, the
spacetime is stationary and non-rotating (i.e.~set
$\Theta=0=\omega_a$), expression (\ref{ddotEa}) reduces to
\begin{eqnarray}
\ddot{E}_a-{\rm D}^2E_a&=&\sigma_{ab}\dot{E}^b-
\sigma_{\langle a}{}^c\sigma_{b\rangle c}E^b+
{\textstyle{4\over3}}\sigma^2E_a+
\varepsilon_{abc}H_d{\rm D}^b\sigma^{cd}-
2\varepsilon_{abc}\sigma^b{}_d{\rm D}^{\langle
c}H^{d\rangle}\nonumber\\&{}& -{\cal R}_{ab}E^b- E_{ab}E^b+
H_{ab}H^b\,,  \label{vddotEa1}
\end{eqnarray}
with an exactly analogous wave equation for $H_a$. When the shear
and the Weyl components are divergence-free (i.e.~for ${\rm
D}^b\sigma_{ab}=0={\rm D}^bE_{ab}={\rm D}^bH_{ab}$), the above
describes the propagation of electromagnetic radiation in the
presence of gravitational waves alone. Thus, using
Eqs.~(\ref{vddotEa1}) one can look form a different perspective
into the age old problem of the interaction between
electromagnetic and gravitational waves in isolated astrophysical
environments, away from the gravitational field of massive compact
stars (e.g.~see~\cite{Co}-\cite{AG} and references therein). In
what follows, however, we will consider a cosmological application
of (\ref{ddotEa}) and (\ref{ddotHa}).

\section{Electromagnetic fields in curved FRW models}
The generic anisotropy of the electromagnetic energy-momentum
tensor makes the Maxwell field incompatible with the high symmetry
of the FRW spacetime. The implication is that the simplest models
where one can study cosmological electromagnetic fields are the
perturbed Friedmann universes.

\subsection{The linear wave equations}
Consider a FRW background cosmology with curved spatial sections.
In covariant terms, the isotropy of the FRW model translates into
$\omega_a=0=\sigma_{ab}=\dot{u}_a$ and $E_{ab}=0=H_{ab}$, while
their spatial homogeneity ensures that all orthogonally projected
gradients vanish identically (i.e.~${\rm D}_a\mu=0={\rm D}_ap={\rm
D}_a\Theta$). This means that $\mu$, $p$, $\Theta$, ${\cal
R}_{ab}={\cal R}h_{ab}/3$ and their time derivatives are the only
non vanishing background quantities.

When studying cosmological electromagnetic fields there is a
widespread perception that, given the conformal invariance of the
Maxwell field and the conformal flatness of the FRW spacetimes,
flat spaces provide an adequate background (e.g.~see~\cite{W,Gi}).
This is only approximately true however, since the FRW symmetries
are generally incompatible with the presence of electric or
magnetic fields. As it is clearly stated in~\cite{SB}, adopting
the conformal triviality of Maxwell's equations on FRW backgrounds
means ignoring the electromagnetic impact on the FRW symmetries.
This is a good approximation when dealing with weak
electromagnetic fields, but only on small scales in models with
nontrivial spatial geometry. In the latter case, the approximation
becomes progressively less accurate as one moves on to larger
scales and the 3-curvature effects start kicking in. Putting it in
another way, with the exception of fully incoherent radiation, one
must study cosmological electromagnetic fields in perturbed
Friedmann universes. The latter, however, are no longer
conformally flat.

On these grounds, we consider a perturbed Friedmann universe with
non-Euclidean spatial sections and allow for the presence a weak
electromagnetic field. The latter vanishes in the background, thus
guaranteeing that both the electric and the magnetic field vectors
are first-order, gauge-invariant perturbations~\cite{SW}. Then,
the source-free components of the nonlinear wave equations
(\ref{ddotEa}) and (\ref{ddotHa}) linearise to
\begin{equation}
\ddot{E}_a- {\rm D}^2E_a=-{\textstyle{5\over3}}\Theta\dot{E}_a-
{\textstyle{4\over9}}\Theta^2E_a+
{\textstyle{1\over3}}\mu(1+3w)E_a- {\cal R}_{ab}E^b\,,
\label{FddotEa}
\end{equation}
and
\begin{equation}
\ddot{H}_a- {\rm D}^2H_a=-{\textstyle{5\over3}}\Theta\dot{H}_a-
{\textstyle{4\over9}}\Theta^2 H_a+
{\textstyle{1\over3}}\mu(1+3w)H_a- {\cal R}_{ab}H^b\,,
\label{FddotHa}
\end{equation}
respectively. During linearisation quantities with nonzero
background value have zero perturbative order, while those that
vanish in the background are first order perturbations and higher
order terms are neglected. For example, the Weyl-free nature of
the FRW metric guarantees that the Weyl effects are nonlinear. The
3-Ricci curvature, on the other hand, contributes to both
(\ref{FddotEa}) and (\ref{FddotHa}). Recall that ${\cal
R}_{ab}=(2k/a^2)h_{ab}$ to zero order, where $k=0,\,\pm1$ is the
curvature index and $a$ represents the scale-factor of the
unperturbed model. In other words, the symmetries of the FRW
metrics ensure that, to linear order, the electromagnetic field
interacts only with the 3-Ricci part of the spacetime curvature.
The curvature terms in (\ref{FddotEa}) and (\ref{FddotHa}) reflect
the earlier mentioned coupling between electromagnetism and
spacetime geometry. Unless the background model is spatially flat,
these are clearly first-order perturbative terms and should be
taken into account in any complete linear study of cosmological
electromagnetic fields. These linear curvature terms clearly show
why large-scale electromagnetic fields are not adequately treated
on flat FRW backgrounds.

Given that the source-free $E_a$ and $H_a$ fields satisfy
identical linear wave equations, we will only consider the
magnetic component and proceed by introducing the following
harmonic decomposition for $H_a$
\begin{equation}
H_a=\sum_{\rm n}H_{({\rm n})}Q_a^{({\rm n})}\,,  \label{Ha}
\end{equation}
where ${\rm n}$ is the comoving eigenvalue of the ${\rm n}$-th
harmonic component and $Q_a^{({\rm n})}$ are the associated vector
harmonics. As usual ${\rm D}_aH^{({\rm n})}=0=\dot{Q}_a^{({\rm
n})}$ and $Q_a^{({\rm n})}$ are eigenfunctions of the
Laplace-Beltrami operator so that ${\rm D}^2Q_a^{({\rm n})}=-({\rm
n}^2/a^2)Q_a^{({\rm n})}$. Employing decomposition (\ref{Ha}) and
introducing the conformal time variable $\eta$ (with
$\dot{\eta}=1/a$) we recast Eq.~(\ref{FddotHa}) as
\begin{equation}
H_{({\rm n})}^{\prime\prime}+{\rm n}^2H_{({\rm n})}=
-4\left(\frac{a^{\prime}}{a}\right)H_{({\rm n})}^{\prime}-
2\left(\frac{a^{\prime}}{a}\right)^2H_{({\rm n})}-
2\left(\frac{a^{\prime\prime}}{a}\right)H_{({\rm n})}- 2kH_{({\rm
n})}\,,  \label{FHa''}
\end{equation}
where a prime indicates differentiation with respect to $\eta$.
Then, on introducing the ``magnetic flux'' variable ${\cal
H}_{({\rm n})}=a^2H_{({\rm n})}$, the above reduces to
\begin{equation}
{\cal H}_{({\rm n})}^{\prime\prime}+n^2{\cal H}_{({\rm n})}=
-2k{\cal H}_{({\rm n})}\,.  \label{FHn''}
\end{equation}
This is a wave equation for ${\cal H}_{({\rm n})}$ with a driving
term on the right-hand side which depends on the background
spatial curvature and vanishes only when the background is
spatially flat. Note that in a model with closed spatial sections
the Laplacian eigenvalue is given by ${\rm n}^2=\nu(\nu+1)$, where
$\nu$ takes the discrete values $\nu=1,2,3,\dots$. Alternatively,
${\rm n}^2=\nu^2+1$ when $k=-1$ and ${\rm n}^2=\nu^2$ for $k=0$
(with $\nu^2\geq0$ in both cases).

\subsection{The linear solutions}
The driving term in the right-hand side of Eq.~(\ref{FHn''}) is
clearly sensitive to the sign of the background spatial curvature.
Let us consider first a FRW model with closed spatial sections.
When $k=+1$, Eq.~(\ref{FHn''}) takes the form
\begin{equation}
{\cal H}_{(\nu)}^{\prime\prime}+ \left[2+\nu(\nu+1)\right]{\cal
H}_{(\nu)}=0\,,  \label{FHn''+}
\end{equation}
with $\nu=1,2,3,\dots$. The above leads to the following
oscillatory solution for the $\nu$-th magnetic mode
\begin{equation}
H_{(\nu)}=\frac{1}{a^2}\left\{{\cal
C}_1\cos\left[\sqrt{2+\nu(\nu+1)}\eta\right] +{\cal
C}_2\sin\left[\sqrt{2+\nu(\nu+1)}\eta\right]\right\}\,,
\label{FHn+}
\end{equation}
where ${\cal C}_1$ and ${\cal C}_2$ are constants. In other words,
for $k=+1$, the magneto-curvature term in the right-hand side of
(\ref{FHn''}) does not have any significant effect on the
evolution of the field, which oscillates in time with an amplitude
that decays according to the $a^{-2}$-law. The only difference
relative to the $k=0$ case is a change in the oscillation
frequency near the long wavelength limit. Note that the
oscillatory behaviour of the field is ensured on all scales by the
compactness of the closed space.

When dealing with the hyperbolic geometry of the spatially open
FRW model, however, the oscillatory behaviour of ${\cal H}_{({\rm
n})}$ is not always guaranteed. Indeed, for $k=-1$
Eq.~(\ref{FHn''}) takes the form
\begin{equation}
{\cal H}_{(\nu)}^{\prime\prime}+ \left(\nu^2-1\right){\cal
H}_{(\nu)}=0\,,  \label{FHn''-}
\end{equation}
with $\nu^2\geq0$. Clearly, when $\nu^2>1$ the harmonic mode
${\cal H}_{(\nu)}$ oscillates just like in a perturbed closed FRW
model. On these scales the background geometry makes no real
difference in the evolution of the field. This agrees with our
perception that curvature effects become progressively less
important as we move towards smaller scales. On sufficiently long
wavelengths (i.e.~for $\nu^2<1$), the geometrical effects take
over and Eq.~(\ref{FHn''-}) no longer accepts an oscillatory
solution. In particular, as $\nu^2\rightarrow0$ we have
\begin{equation}
{\cal H}_{(\nu)}= {\cal C}_1\cosh\eta+ {\cal C}_2\sinh\eta= {\cal
C}_3e^{\eta}+{\cal C}_4e^{-\eta}\,,  \label{FHn}
\end{equation}
where ${\cal C}_1$ and ${\cal C}_2$ are constants and ${\cal
C}_{3,4}=({\cal C}_1\pm{\cal C}_2)/2$. Note that, since
$n^2=\nu^2+1>1$ always, these long wavelength solutions still
correspond to subcurvature modes~\cite{LW}. To have a closer look
at the effect of geometry on the linear evolution of the field, we
note that the evolution of a spatially open FRW model is monitored
by
\begin{equation}
a\Theta=3\coth(\beta\eta)\,,  \label{eta}
\end{equation}
with $\beta=(1+3w)/2$ by definition and $\beta\eta>0$
(e.g.~see~\cite{C}). The above holds throughout the various
periods in the lifetime of an open FRW universe, provided the
barotropic index $w$ remains constant during each epoch. Then, the
relation between the scale factor and the conformal time variable
is
\begin{equation}
a=a_0\left(\frac{1-e^{-2\beta\eta}}{1-e^{-2\beta\eta_0}}\right)^{1/\beta}
e^{\eta-\eta_0}\,,  \label{a}
\end{equation}
where $\eta_0$, $a_0$ depend on the initial conditions. Throughout
the dust era $w=0$ and $\beta=1/2$, while $w=1/3$ and $\beta=1$
when radiation dominates. Finally, during a period of inflationary
expansion with $p=-\rho$ we have $\beta=-1$. Note that in the
latter case the conformal time variable takes negative values.
According to expression (\ref{a}), there are extensive periods in
the lifetime of the universe (i.e.~as long as $\eta\ll0$ or
$\eta\gg0$) when the relation between the cosmological scale
factor and the conformal time variable is (see also~\cite{Ba})
\begin{equation}
a\propto e^{\eta}\,.  \label{a1}
\end{equation}
Substituting this result into the right-hand side of
Eq.~(\ref{FHn}), and taking into account that ${\cal
H}_{({\nu})}=a^2H_{({\nu})}$ by definition, we arrive at
\begin{equation}
H_{(\nu)}={\cal C}_3a^{-1}+ {\cal C}_4a^{-3}\,.  \label{Hn}
\end{equation}
Therefore, large-scale magnetic fields in perturbed spatially open
FRW models decay as $a^{-1}$, a rate considerably slower than the
standard ``adiabatic'' $a^{-2}$-law. The immediate consequence is
that, at the long wavelength limit, the cosmological magnetic flux
is no longer conserved. Instead, the product $a^2H_{(\nu)}$
increases with time. This opens the possibility of an effective
superadiabatic amplification of the field on large scales similar
to that found in~\cite{TW}. Even if the universe is only
marginally open today, this effect could have important
implications for the present strength of primordial large-scale
magnetic fields. Particularly for those fields that survived an
epoch of inflation, since they would be much stronger than
previously anticipated. Note that during inflation the
conductivity of the cosmic medium is effectively zero, which in
turn ensures the absence of spatial currents (see Section 3.4). In
this article, we have focused primarily on the mathematics of the
magneto-geometrical interaction and provided a qualitative measure
of its implications for large-scale magnetic fields. A discussion
of the physics, together with a detailed quantitative study of the
amplification effect, will be given in a subsequent article.

So far, similar modifications in the evolution of cosmological
magnetic fields have been obtained at the expense of standard
electromagnetic properties, and in particular of the conformal
invariance of the Maxwell's equations
(e.g.~see~\cite{TW}-\cite{AM} for a representative, though
incomplete, list). Moreover, in some cases this effect is achieved
by introducing new physics. Our analysis shows that one can still
arrive at the same result by taking into account the natural,
general relativistic coupling between the electromagnetic field
and the spacetime curvature. In other words, contrary to the
widespread perception, superadiabatic magnetic amplification is
possible within conventional electromagnetic theory. Here, this
has been done through the field's coupling to the intrinsic
curvature of spatially open FRW models. Interestingly, however,
analogous effects can also occur in perturbed flat FRW cosmologies
by coupling the magnetic field to the Weyl curvature of the model,
namely to the gravitational waves~\cite{TDM}. All these cast new
light on the role and the potential implications of spacetime
geometry for the evolution of large-scale cosmic magnetic fields.

\section{Discussion}
The general relativistic coupling between the electromagnetic and
the gravitational fields has long been known in the literature. So
far, this interaction has been primarily studied in terms of the
Faraday tensor and of the electromagnetic
4-potential~\cite{DWB}-\cite{N}. Here, we have taken an
alternative approach by looking at the evolution of the individual
electromagnetic field components in a general curved spacetime.
Assuming that the matter field is of the perfect fluid form, we
have derived from first principle the nonlinear wave equations of
the electric and the magnetic parts of the Maxwell field. This
complements earlier studies which have provided a
differential/integral formulation of Maxwell's formulae in terms
of the physically measurable components of the electromagnetic
field (e.g.~see~\cite{TM}-\cite{H}). Our approach identifies and
isolates all the sources that drive the propagation of
electromagnetic fields by keeping the separate aspects of the
problem quite distinct. Also, by being manifestly covariant at
every step, our calculation avoids undue complexity without
introducing any specific coordinate frame. We show explicitly how
the electric and magnetic fields are affected by the various
kinematical and dynamical quantities and particularly by the
different parts of the gravitational field.

Given that large-scale electromagnetic fields are generally
incompatible with the FRW symmetries, we consider perturbed models
and concentrate on the evolution of large-scale magnetic fields.
In particular, we linearise our equations about spatially curved
FRW spacetimes and investigate the implications of the background
curvature for the evolution of cosmological electromagnetic
fields. The gauge-invariance of our linear equations ensures that
our results are free from any gauge-related problems and
ambiguities. We show that when the zero-order spacetime has open
spatial sections, the magnetic flux is not always conserved. More
specifically, magnetic fields coherent on the largest subcurvature
scales are found to decay as $a^{-1}$, instead of following the
familiar $a^{-2}$-law, where $a$ is the cosmological scale factor.
The reason for this deviation is the general relativistic coupling
between the magnetic field and the intrinsic curvature of a
perturbed spatially open FRW universe. This magneto-geometrical
interaction can change the evolution of the field on large scales,
where curvature effects become important. The result is a natural
superadiabatic-type amplification of cosmological magnetic fields,
without the need for new physics and without breaking away from
standard electromagnetism.

\section*{Acknowledgments}
The author would like to thank John Barrow, Naresh Dadhich, Nader
Haghighipour, Brian Pitts and John Stewart for helpful discussions
and comments.

\end{document}